\documentclass[aps,prl,twocolumn,superscriptaddress]{revtex4-1}

\usepackage{graphicx}
\usepackage{float}
\usepackage{multirow}
\usepackage{lineno}

\newcommand{\pt}{$p_{T}$ }

\newcommand{\hla}{$\mathrm{^{3}_{\Lambda}H}$ }
\newcommand{\ahla}{$\mathrm{^{3}_{\overline{\Lambda}}\overline{H}}$ }
\newcommand{\sNN}{$\sqrt{s_{_\mathrm{NN}}}$ }
\newcommand{\he}{$\mathrm{^{3}He}$ }
\newcommand{\ahe}{$\mathrm{^{3}\overline{He}}$ }

\newcommand{\lifetime}          {$142^{+24}_{-21}\,{\rm (stat.)} {\pm} 31\,{\rm (syst.)} $ }

\newcommand{\lifetimeNoSys}{$142^{+24}_{-21} $ }

\newcommand{\channelThree} {$^3_\Lambda$H $\rightarrow d + p + \pi^-$ }
\newcommand{\dlbg}               {$ \ell / \beta \gamma$ }  

\newcommand{\ratio}               {$0.32\rm{\pm}0.05\,{\rm (stat.)}\pm 0.08\,{\rm (syst.)}$}

\newcommand{\lifetimeJH}      {$182^{+89}_{-45}\,{\rm (stat.)}\pm 27\,{\rm (syst.)}$ }

\begin{document}

\title{Measurement of \hla lifetime in Au+Au collisions at the Relativistic Heavy-Ion Collider}

\affiliation{AGH University of Science and Technology, FPACS, Cracow 30-059, Poland}
\affiliation{Argonne National Laboratory, Argonne, Illinois 60439}
\affiliation{Brookhaven National Laboratory, Upton, New York 11973}
\affiliation{University of California, Berkeley, California 94720}
\affiliation{University of California, Davis, California 95616}
\affiliation{University of California, Los Angeles, California 90095}
\affiliation{Central China Normal University, Wuhan, Hubei 430079}
\affiliation{University of Illinois at Chicago, Chicago, Illinois 60607}
\affiliation{Creighton University, Omaha, Nebraska 68178}
\affiliation{Czech Technical University in Prague, FNSPE, Prague, 115 19, Czech Republic}
\affiliation{Nuclear Physics Institute AS CR, 250 68 Prague, Czech Republic}
\affiliation{Frankfurt Institute for Advanced Studies FIAS, Frankfurt 60438, Germany}
\affiliation{Institute of Physics, Bhubaneswar 751005, India}
\affiliation{Indiana University, Bloomington, Indiana 47408}
\affiliation{Alikhanov Institute for Theoretical and Experimental Physics, Moscow 117218, Russia}
\affiliation{University of Jammu, Jammu 180001, India}
\affiliation{Joint Institute for Nuclear Research, Dubna, 141 980, Russia}
\affiliation{Kent State University, Kent, Ohio 44242}
\affiliation{University of Kentucky, Lexington, Kentucky 40506-0055}
\affiliation{Lamar University, Physics Department, Beaumont, Texas 77710}
\affiliation{Institute of Modern Physics, Chinese Academy of Sciences, Lanzhou, Gansu 730000}
\affiliation{Lawrence Berkeley National Laboratory, Berkeley, California 94720}
\affiliation{Lehigh University, Bethlehem, Pennsylvania 18015}
\affiliation{Max-Planck-Institut fur Physik, Munich 80805, Germany}
\affiliation{Michigan State University, East Lansing, Michigan 48824}
\affiliation{National Research Nuclear University MEPhI, Moscow 115409, Russia}
\affiliation{National Institute of Science Education and Research, HBNI, Jatni 752050, India}
\affiliation{National Cheng Kung University, Tainan 70101 }
\affiliation{Ohio State University, Columbus, Ohio 43210}
\affiliation{Institute of Nuclear Physics PAN, Cracow 31-342, Poland}
\affiliation{Panjab University, Chandigarh 160014, India}
\affiliation{Pennsylvania State University, University Park, Pennsylvania 16802}
\affiliation{Institute of High Energy Physics, Protvino 142281, Russia}
\affiliation{Purdue University, West Lafayette, Indiana 47907}
\affiliation{Pusan National University, Pusan 46241, Korea}
\affiliation{Rice University, Houston, Texas 77251}
\affiliation{Rutgers University, Piscataway, New Jersey 08854}
\affiliation{Universidade de Sao Paulo, Sao Paulo, Brazil, 05314-970}
\affiliation{University of Science and Technology of China, Hefei, Anhui 230026}
\affiliation{Shandong University, Jinan, Shandong 250100}
\affiliation{Shanghai Institute of Applied Physics, Chinese Academy of Sciences, Shanghai 201800}
\affiliation{State University of New York, Stony Brook, New York 11794}
\affiliation{Temple University, Philadelphia, Pennsylvania 19122}
\affiliation{Texas A\&M University, College Station, Texas 77843}
\affiliation{University of Texas, Austin, Texas 78712}
\affiliation{University of Houston, Houston, Texas 77204}
\affiliation{Tsinghua University, Beijing 100084}
\affiliation{University of Tsukuba, Tsukuba, Ibaraki, Japan,305-8571}
\affiliation{Southern Connecticut State University, New Haven, Connecticut 06515}
\affiliation{University of California, Riverside, California 92521}
\affiliation{University of Heidelberg}
\affiliation{United States Naval Academy, Annapolis, Maryland 21402}
\affiliation{Valparaiso University, Valparaiso, Indiana 46383}
\affiliation{Variable Energy Cyclotron Centre, Kolkata 700064, India}
\affiliation{Warsaw University of Technology, Warsaw 00-661, Poland}
\affiliation{Wayne State University, Detroit, Michigan 48201}
\affiliation{World Laboratory for Cosmology and Particle Physics (WLCAPP), Cairo 11571, Egypt}
\affiliation{Yale University, New Haven, Connecticut 06520}

\author{L.~Adamczyk}\affiliation{AGH University of Science and Technology, FPACS, Cracow 30-059, Poland}
\author{J.~R.~Adams}\affiliation{Ohio State University, Columbus, Ohio 43210}
\author{J.~K.~Adkins}\affiliation{University of Kentucky, Lexington, Kentucky 40506-0055}
\author{G.~Agakishiev}\affiliation{Joint Institute for Nuclear Research, Dubna, 141 980, Russia}
\author{M.~M.~Aggarwal}\affiliation{Panjab University, Chandigarh 160014, India}
\author{Z.~Ahammed}\affiliation{Variable Energy Cyclotron Centre, Kolkata 700064, India}
\author{N.~N.~Ajitanand}\affiliation{State University of New York, Stony Brook, New York 11794}
\author{I.~Alekseev}\affiliation{Alikhanov Institute for Theoretical and Experimental Physics, Moscow 117218, Russia}\affiliation{National Research Nuclear University MEPhI, Moscow 115409, Russia}
\author{J.~Alford}\affiliation{Kent State University, Kent, Ohio 44242}
\author{D.~M.~Anderson}\affiliation{Texas A\&M University, College Station, Texas 77843}
% ...

\collaboration{STAR Collaboration}\noaffiliation

\date{\today}

\begin{abstract}
A precise measurement of the \hla lifetime is presented. In this letter, the mesonic decay modes $\mathrm{{^3_\Lambda}H\rightarrow}$\he + $\pi^-$ and \channelThree are used to reconstruct the \hla from Au+Au collision data collected by the STAR collaboration at RHIC. A minimum $\chi^2$ estimation is used to determine the lifetime of $\tau = $ \lifetime ps. This lifetime is about 50\% shorter than the lifetime $\tau = 263\pm2$ ps of a free $\Lambda$, indicating strong hyperon-nucleon interaction in the hypernucleus system. The branching ratios of the mesonic decay channels are also determined to satisfy B.R.$_{(^3{\rm He}+\pi^-)}/$(B.R.$_{(^3{\rm He}+\pi^-)}+$B.R.$_{(d+p+\pi^-)}) = $ \ratio. Our ratio result favors the assignment $J(\mathrm{^{3}_{\Lambda}H})$ = $\frac{1}{2}$ over $J(\mathrm{^{3}_{\Lambda}H})$ = $\frac{3}{2}$. These measurements will help to constrain models of hyperon-baryon interactions.

\end{abstract}


\pacs{21.80.+a; 21.10.Tg; 21.10.Dr}

\maketitle


The hyperon-nucleon (Y-N) interaction is of fundamental interest because it introduces the strangeness quantum number in nuclear matter~\cite{2012.EPJA.48.41} and so understanding it can provide insights into the strong interaction, often through the use of effective models that extend work on normal nuclei to the flavor SU(3) group~\cite{2000.JPG.26.1049}. The Y-N interaction is also of crucial importance in high-density matter systems, such as neutron stars~\cite{2004.Science.304.536,2008.NPA.804.309}. At such high densities, particles with some strange content can be created. The formation of hyperons softens the equation of state and reduces the possible maximum mass of the corresponding neutron star~\cite{Glendenning:1997wn}, which makes it extremely difficult to describe neutron stars exceeding two solar masses, such as those observed recently in~\cite{2010.Nature.467.1081,2013.Science.340.448}. Among other explanations (such as deconfinement to quark matter), alternative Y-N couplings have been suggested as possible solutions for the so-called ``hyperon puzzle"~\cite{2012.PRC.85.065802,2014.PRC.89.025805,2015.AstroPhysJ.808.8}. 

Hypernuclei are natural hyperon-baryon correlation systems and can be used as an experimental probe to study the Y-N interaction. The lifetime of a hypernucleus depends on the strength of the Y-N interaction \cite{1962.PL.1.58,1998.PRC.57.1595}. Therefore, a precise determination of the lifetime of hypernuclei provides direct information on the Y-N interaction strength \cite{1998.PRC.57.1595,1973.NPB.52.1}.

The hypertriton $\mathrm{^{3}_{\Lambda}H}$, which consists of a $\Lambda$, a proton and a neutron, is the lightest known hypernucleus. It has been argued that if the \hla is a $\Lambda$ hyperon weakly bound to a deuteron core, then the lifetime of the \hla should be close to that of the free $\Lambda$ \cite{1966.ILNC.46.786}. The lifetime of the \hla has been measured using helium bubble chambers and nuclear emulsion since the 1960s \cite{1964.Proc.ICHF63.63,1964.PR.136.B1803,1968.PRL.20.819,1969.PR.180.1307,1970.NPB.16.46,1970.PRD.1.66,1973.NPB.67.269}. The first measurement from a helium bubble chamber experiment yielded $\tau$($\mathrm{^{3}_{\Lambda}H}$) = $95^{+ 19}_{- 15}$ ps \cite{1964.Proc.ICHF63.63}. Subsequent measurements indicated a lifetime close to \cite{1968.PRL.20.819,1969.PR.180.1307,1970.PRD.1.66,1973.NPB.67.269} or shorter than \cite{1964.PR.136.B1803,1970.NPB.16.46} that of the free $\Lambda$, though with large statistical uncertainty. Recent measurements of the \hla lifetime from experiments at RHIC (BNL), HypHI (GSI) and LHC (CERN) were reported \cite{2010.Science.328.58,2013.NPA.913.170,2016.PLB.754.360}. They all show a lifetime shorter than that of the free $\Lambda$. However, due to the dispersion of the different measurements, a clear conclusion on the lifetime of \hla cannot be reached. Moreover, theoretical calculations do not provide a consensus picture of \hla structure because of the diverging lifetime values \cite{1959.PR.116.1312,1962.PL.1.58,1963.PRL.11.96,1966.ILNC.46.786,1971.NPB.28.566,1979.ILNC.51.180,1988.SPJ.31.210,1992.JPG.18.339,1998.PRC.57.1595}.

In this letter, we report a new precise measurement of the \hla lifetime from the STAR (Solenoid Tracker at RHIC) experiment. RHIC provides an ideal laboratory to study the Y-N interaction because hyperons and nucleons are abundantly produced in high-energy nucleus-nucleus collisions \cite{2010.Science.328.58}. The main detector of STAR \cite{2003.NIMA.499.624} is a time projection chamber (TPC) \cite{2003.NIMA.499.659} that measures momentum and energy loss of particles produced in heavy-ion collisions. This information is used to identify charged particles, like $\pi^\pm$, $p$, $d$ and \he produced in the collisions. We are able to reconstruct \hla via its two main decay channels: $\mathrm{{^3_\Lambda}H\rightarrow}$\he + $\pi^-$ and {$^3_\Lambda$H $\rightarrow d + p + \pi^-$.} The theoretical branching ratios for those two channels are 24.89\% and 40.06\%, respectively~\cite{1998.PRC.57.1595}. Due to small branching ratios, or decays into neutral particles \cite{1998.PRC.57.1595}, the remaining decay channels have been disregarded in this paper. 

The beam energy scan at RHIC during the years 2010 and 2011 allowed STAR to collect data from Au+Au collisions over a broad range of energies. The lifetime is an intrinsic property of every unstable particle, and is independent of beam energy \cite{2014.CPC.38.090001}. All \hla measurements, regardless of beam energy, are combined to increase the statistics.

A minimum-bias (MB) trigger at multiple beam energies was used. For the 2-body decay channel analysis, we use data from six different energies, \sNN = 7.7, 11.5, 19.6, 27, 39, and 200 GeV; for the 3-body decay analysis, we have three beam energies, \sNN = 27, 39, and 200 GeV. The 200 GeV data used in the 2-body analysis were collected in 2010, and data for the 3-body channel were collected in 2011. The current paper includes a 2-body decay analysis that was completed prior to the availability of newer samples~\cite{Yuhui-thesis}. As a cross-check, a 3-body decay analysis was subsequently carried out; this was confined to 2011 datasets which offered better statistics and lower backgrounds for that channel~\cite{Yifei-thesis}. Nevertheless, we report results that represent substantial improvements in statistical uncertainties over prior measurements. Further improvements in \hla measurements are expected when future runs become available for analysis. The event statistics and basic event-level selections for the 2-body and the 3-body channel analyses are listed in Tables \ref{Tab.2Body} and \ref{Tab.3Body}, respectively. In addition, the counts of well identified \he and \ahe candidates are listed for the 2-body decay mode in Table \ref{Tab.2Body}. The numbers of identified \hla and \ahla are listed in Table \ref{Tab.2Body} and only identified \hla are listed in Table \ref{Tab.3Body}. 

\begin{figure}[htb]
\centering
\includegraphics[width=0.23\textwidth]{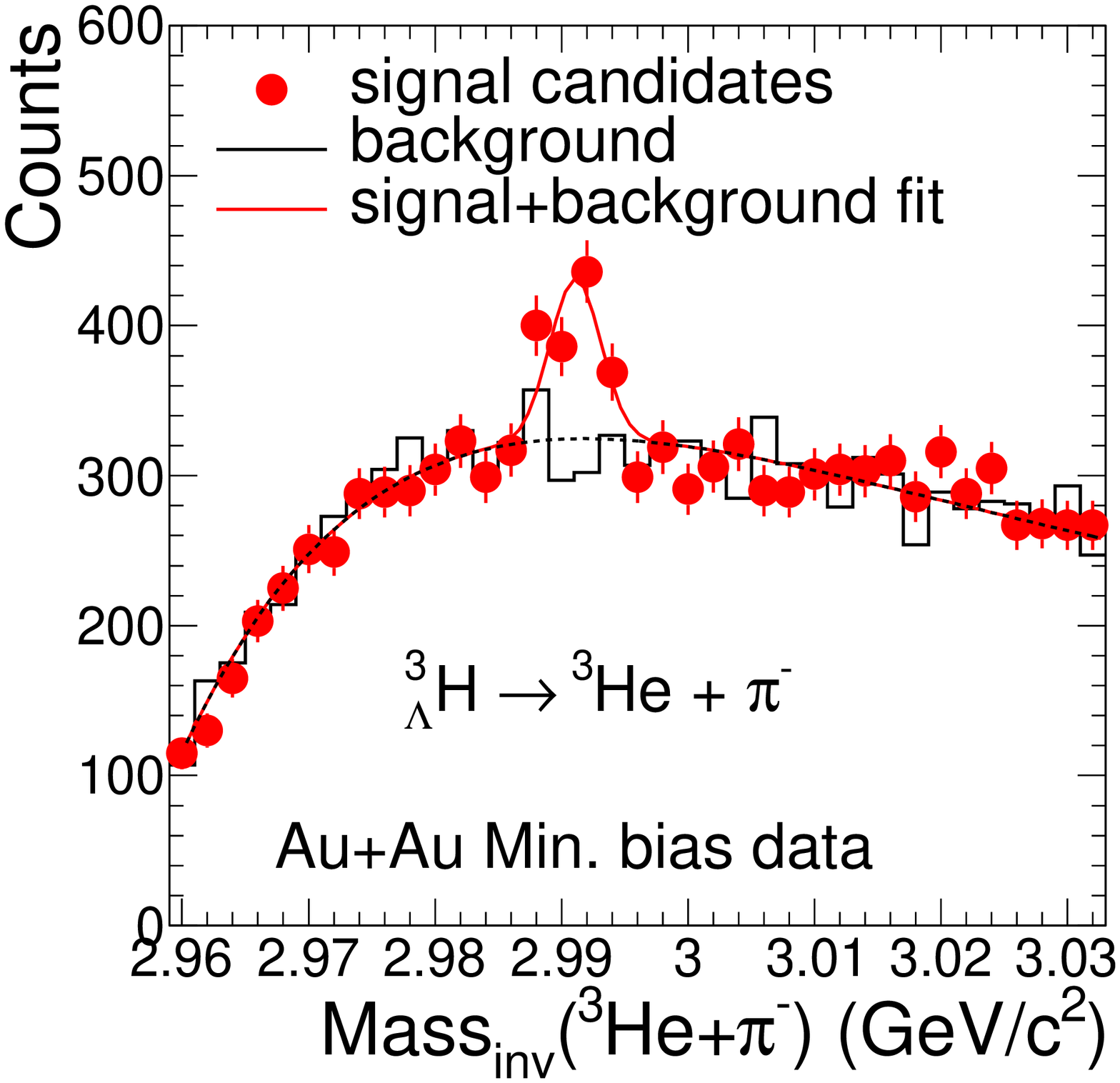}
\includegraphics[width=0.23\textwidth]{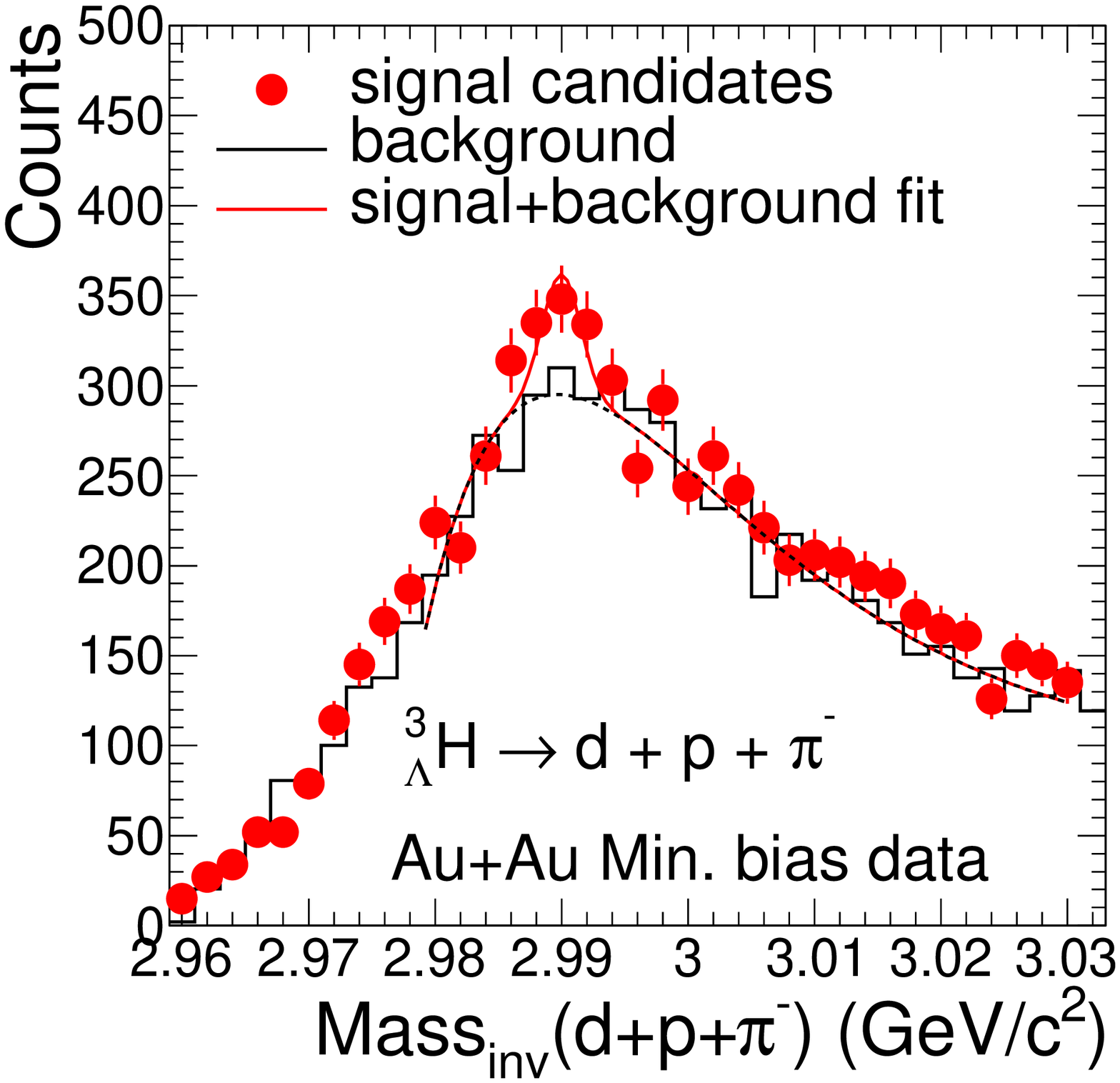}
\caption{(Color Online) The \hla invariant mass distribution for each decay channel. The solid circles represent the signal candidate distributions, and the solid histograms are the rotated background. The background shapes were constrained by fits, shown as dotted black lines. The solid red lines are a fit combining signal (Gaussian) plus background (double exponential). Error bars represent statistical errors.
\label{Fig.Inv}}
\end{figure}

\begin{table}[htbp]
\centering
\caption{Dataset for the 2-body decay channel analysis, with \he and \hla statistics.}
\label{Tab.2Body}
\begin{tabular}{lccccc}
\hline
Energy & Events ($\times$ 10M)& \he & \ahe & \hla+\ahla \\
\hline \hline

7.7 GeV   &    ~0.4  & 6388$\pm$80 &       0                  & 52$\pm$17 \\ 
11.5 GeV &   ~1 & 5330$\pm$73 &       0                  & 44$\pm$16 \\
19.6 GeV &   ~3 & 4941$\pm$70 &       0                 & 42$\pm$14 \\
27 GeV    &   ~5 & 4179$\pm$65 &     19$\pm$4     & 45$\pm$16 \\
39 GeV    & ~12 & 5252$\pm$72 &   133$\pm$12   & 86$\pm$21 \\
200 GeV  & ~22 & 6850$\pm$83 &  2213$\pm$47  & 85$\pm$20 \\
\hline
\end{tabular}

\end{table}
\begin{table}[htbp]
\centering
\caption{Dataset for the 3-body decay channel analysis, with \hla statistics.}
\label{Tab.3Body}
\begin{tabular}{lcc}
\hline
Energy  & Events ($\times$ 10M)& \hla \\
\hline \hline

27 GeV   & ~5   &  42$\pm$16\\
39 GeV   & ~13 &  53$\pm$13\\
200 GeV & ~52 & 128$\pm$30\\
\hline
\end{tabular}
\end{table}

The \hla candidates are reconstructed from the invariant mass distributions of the daughters: \he + $\pi^-$ for the 2-body decay channel, and $d+p+\pi^-$ for the 3-body decay channel, shown as solid circles in Fig. \ref{Fig.Inv}. Tracks with transverse momentum \pt $> 0.2$ GeV/$c$ and pseudorapidity $|\eta|<1.0$ are used for \hla candidate reconstruction. The \hla has a typical decay length of several centimeters, which is long enough to be resolved by the STAR TPC. To optimize the signal to background ratio, we apply a combination of constraints to the decay topology parameters, including the distance of closest approach (DCA) between daughter tracks, the DCA of daughters to the \hla decay vertex, the DCA of the \hla candidate to the primary heavy-ion collision vertex, the decay length of the \hla candidate, and the DCA of the daughters to the collision vertex. Topology selections are optimized separately for the 2-body and 3-body decay channels, with the selections for the 2-body case being very similar to those listed in the STAR 2010 publication \cite{2010.Science.328.58}. 

Using the candidates that pass the topology selections, a background invariant mass curve is constructed by rotating one of the daughters $180^\circ$ in azimuthal angle. The $\pi^-$ is rotated in the case of the 2-body channel, and the deuteron in the case of the 3-body channel. This procedure accurately describes the residual combinatorial background shown as solid histograms in Fig. \ref{Fig.Inv}. The background shapes are fitted by a double exponential function: $f(x)\propto \exp(-x /p_1) - \exp(-x /p_2)$. The signals are then fitted by adding a Gaussian function to the background. Bin-by-bin counting is used to calculate the signal within the mass range [2.987, 2.995] GeV/$c^2$. In total, 354 and 223 \hla candidates are identified in 2-body and 3-body channel analyses, respectively.

\begin{figure}
\includegraphics[width=0.23\textwidth]{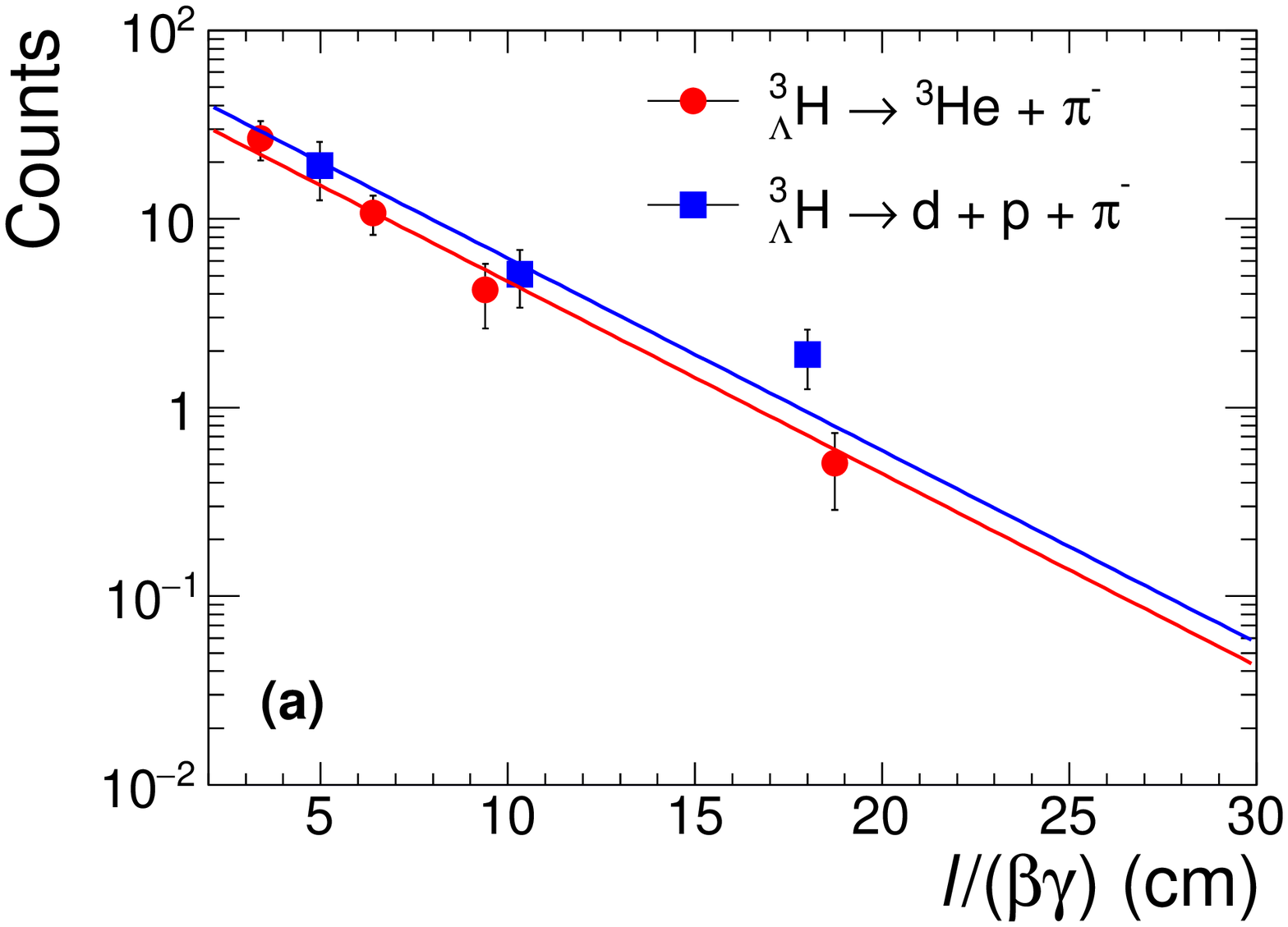}
\includegraphics[width=0.23\textwidth]{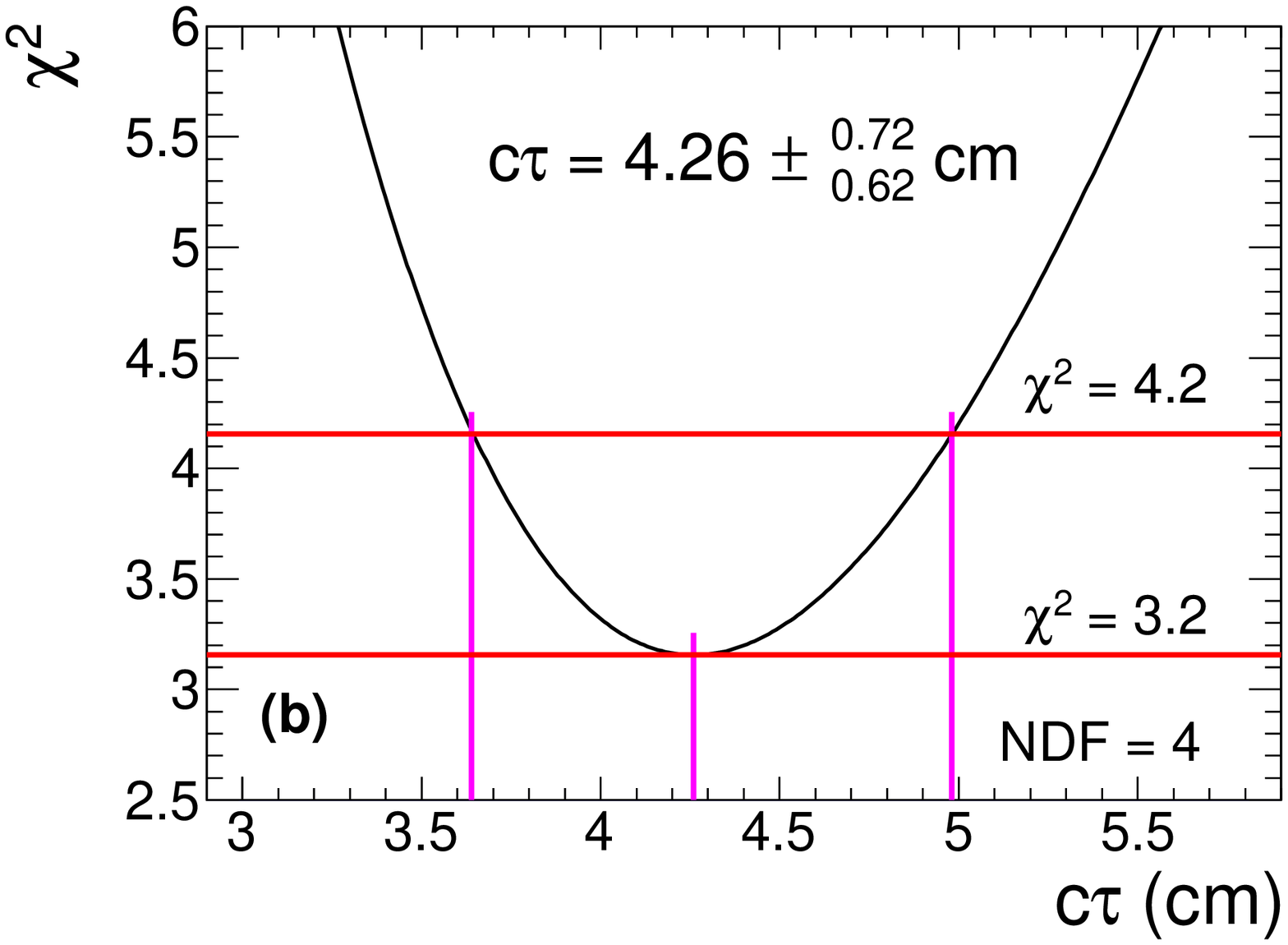}
\caption{(Color Online) Panel (a): The \hla yield as a function of \dlbg for each of the two analyzed decay channels. The red points are for 2-body decays in four bins of \dlbg, and the blue squares are for 3-body decay in three \dlbg bins. The yields indicate the number of \hla per million events for each channel, and are already divided by the theoretical branching ratio (40.06\% for the 3-body channel and 24.89\% for the 2-body channel \cite{1998.PRC.57.1595}). The data points are fitted with the usual radioactive decay function. Panel (b): The best fit result to the seven data points in panel (a) using a minimum $\chi^2$ estimation.}
\label{Fig.lifetime}
\end{figure}

The \hla decays obey  $N(t) = N_0 e^{-t/\tau} = N_0 e^{-\ell / \beta \gamma c\tau} $, where $\ell$ is the \hla decay length, $\beta=v/c$, and $\gamma$ is the Lorentz factor. For the 2-body decay channel, we count \hla decays in four bins of \dlbg: [2, 5] cm, [5, 8] cm, [8, 11] cm, and [11, 41] cm. Because the 3-body decay channel has fewer events due to a lower reconstruction efficiency with a magnitude of 1\%, only three bins in \dlbg are used in this decay channel: [2.4, 8] cm, [8, 13] cm, and [13, 25] cm. We correct the \hla counts in each bin for reconstruction efficiency and detector acceptance using STAR embedding data, which is derived from a Monte-Carlo GEANT3 simulation with STAR detector geometry~\cite{FINE200176}. The yield in each bin is computed according to the number of events used for the 2-body and 3-body analyses by normalizing to \he counts in the same experiment, and the results are shown in panel (a) of Fig. \ref{Fig.lifetime}. The lifetime is extracted from the fit to the \dlbg distribution. Asymmetric statistical errors are calculated by performing a minimum $\chi^2$ estimation of the fit to the $c\tau$ distributions as represented in panel (b) of Fig. \ref{Fig.lifetime}. Our result is \lifetimeNoSys ps, shown as crosses of horizontal and vertical lines in panel (b) of Fig. \ref{Fig.lifetime}. As a comparison, the \hla lifetime measurement reported by STAR in 2010 \cite{2010.Science.328.58} is \lifetimeJH ps. The present measurement is consistent with STAR's 2010 measurement to within 0.9$\sigma$ and has a smaller uncertainty.

Systematic errors fall into several main categories. First, we consider systematics arising from the values chosen for topology cuts. Second, the effect of the choice of bin width for the \hla candidate invariant mass plots was investigated. Third, we investigate systematics due to the properties of \hla assumed in the embedding analysis, by varying both the assumed \pt distribution and assumed lifetime of the $\mathrm{^{3}_{\Lambda}H}$. We also investigated the contribution from comparison with side-band techniques \cite{2010.Science.328.58}. Details of those systematic errors are shown in Table \ref{Tab.sys}. Additional sources of systematics, including loss of \hla due to interactions between \hla and the detector material or gas are found to be negligible. The individual contributions are added in quadrature and are reflected in the final systematic error of 31 ps. 

\begin{table}[htbp]
\centering
\caption{Main sources of systematic uncertainty for lifetime measurement in the 2-body and 3-body decay analyses.}
\begin{tabular}{ccc}
\hline
            Decay channel & Systematic source & Uncertainty(\%) \\
\hline \hline
\multirow{5}{*}{2-body} & Invariant mass binning                    & 5.69 \\
                                     & Decay length and DCA ($\pi$)         & 2.44 \\
                                     & DCA ($^3$He to $\pi$)                     & 5.69 \\
                                     & Embedding analysis			    & 6.50 \\
                                     & Background shape                         & 3.51 \\
\hline
\multirow{5}{*}{3-body} & Invariant mass binning                     &  8.76 \\
                                     & DCA ($p$ to $\pi$)                            &  2.58 \\
                                     & DCA ($p\text{-}\pi$ pair)                   & 14.95 \\
                                     & Embedding analysis			     & 4.93 \\
                                     & Background shape                           & 3.56 \\
\hline
\label{Tab.sys}
\end{tabular}
\end{table}

As a further cross-check, the $\Lambda$ has been reconstructed via the $\Lambda \rightarrow p + \pi^-$ decay channel in our experiment using the same method, and we obtain $267 \pm 5$~ps for the $\Lambda$ lifetime~\cite{2010.Science.328.58}. This measurement is consistent with the $\Lambda$ lifetime of $263 \pm 2$ ps compiled by the Particle Data Group \cite{2014.CPC.38.090001}.

\begin{figure}
\includegraphics[width=0.5\textwidth]{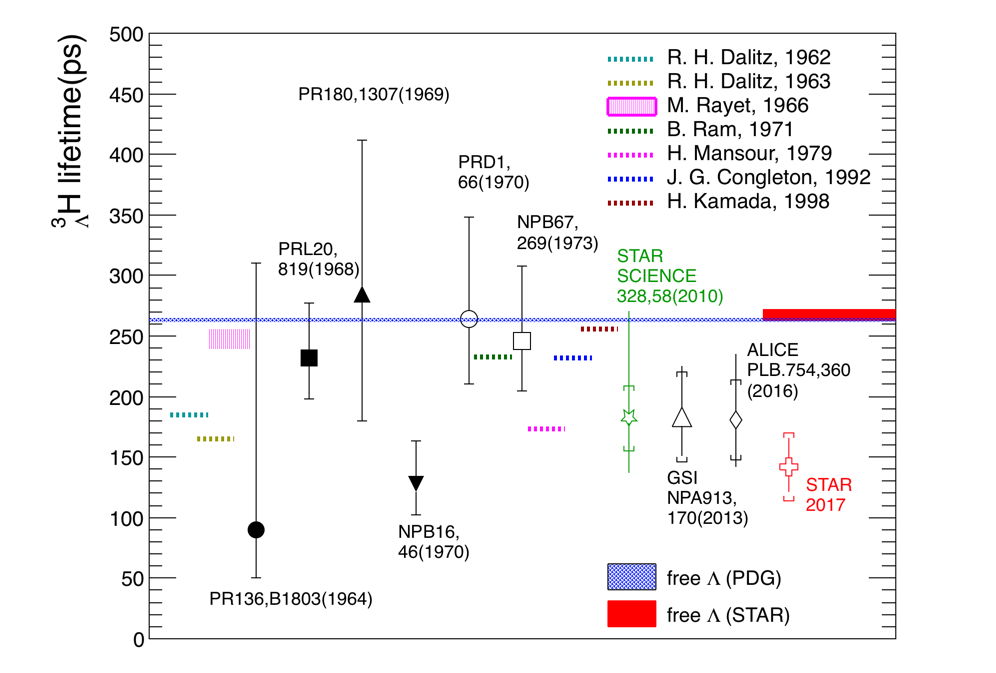}
\caption{(Color Online) A summary of worldwide \hla lifetime experimental measurements and theoretical calculations. The two star markers are the STAR collaboration's measurement published in 2010 \cite{2010.Science.328.58} and the present analysis.}
\label{Fig.Wdata}
\end{figure}

A summary plot of the worldwide \hla lifetime measurements is shown in Fig. \ref{Fig.Wdata}. There have been discussions of the lifetime of \hla since the 1960s. For many years, the \hla was considered as a weakly-bound state formed from a deuteron and a $\Lambda$, which leads to the inference that the \hla lifetime should be very close to that of the free $\Lambda$. However, not all experimental measurements support this picture. From Fig. \ref{Fig.Wdata}, it can be seen that there are at least two early measurements \cite{1964.PR.136.B1803,1970.NPB.16.46} that indicate \hla has a shorter lifetime than the $\Lambda$. The lifetime measured in \cite{1970.NPB.16.46} has the smallest error among similar studies in the 1960s and 70s, and was shorter than the others. This measurement was based on the 3-body decay channel $^3_\Lambda$H $\rightarrow p + d + \pi^-$ in a nuclear emulsion experiment. The shorter lifetime was attributed to the dissociation of the lightly-bound $\Lambda$ and deuteron when traveling in a dense medium. However, this explanation is not fully convincing since measurements in Refs. \cite{1968.PRL.20.819,1969.PR.180.1307,1973.NPB.67.269} also used nuclear emulsion, yet their results were close to the $\Lambda$ lifetime. In addition, Ref. \cite{1964.PR.136.B1803} used a helium bubble chamber that should not be affected by the hypothesized dissociation, and report a lifetime lower than that of the free $\Lambda$.

A recent statistical compilation of the lifetime measurements available in the literature favors the lifetime of \hla ($215^{+18}_{-16}$ ps) being shorter than that of the $\Lambda$ \cite{2014.PLB.728.543,2016.PLB.754.360}. The present lifetime measurement casts further doubt on the early inferences concerning the structure of the $\mathrm{^3_\Lambda H}$. The lifetime is related to the binding energy of the $\Lambda$ in this hypernucleus and to its decay channels. Theoretical predictions need to employ assumptions about the $\Lambda$ binding energy, which is poorly measured \cite{1998.PRC.57.1595,1973.NPB.52.1}. Assuming a larger binding energy leads to a shorter lifetime \cite{1962.PL.1.58}. There is also the possibility that stimulated $\Lambda$-decay due to the presence of other nucleons, such as the process $\mathrm{\Lambda + N \rightarrow N + N + \pi^0}$ may contribute to the pionic modes \cite{1962.PL.1.58}. This effect may become much larger due to interference with the normal decay interaction \cite{1962.PL.1.58}. The current measurements clearly motivate further theoretical study \cite{2016.RMP.88.035004}.

Because the \hla can be reconstructed via its two decay channels, $\mathrm{{^3_\Lambda}H\rightarrow}$\he + $\pi^-$ and \channelThree at STAR, it is possible to compare the decay branching ratios for those two channels. By fitting the data points in Fig. \ref{Fig.lifetime}(a) with the radioactive decay function, we can extract the product $N_0\times$B.R. for each channel. We define 
$${\rm Ratio} = \frac{\mathrm{B.R.(^3_\Lambda H \rightarrow~^3He + \pi^-)}} {\mathrm{B.R.(^3_\Lambda H \rightarrow~^3He + \pi^-)} + \mathrm{B.R.(^3_\Lambda H \rightarrow d + p + \pi^-)} }$$
This definition is different from a more commonly used variable, $R_3$, which is defined as:
$${R_3} = \frac{\mathrm{B.R.(^3_\Lambda H \rightarrow~^3He + \pi^-)}}{\mathrm{B.R.(^3_\Lambda H \rightarrow all~\pi^-~channels)}}$$
however, considering that, theoretically, the sum of B.R.s of $\mathrm{{^3_\Lambda}H\rightarrow}$\he + $\pi^-$ and \channelThree channels is over 99\% of all $\pi^-$ channels~\cite{1998.PRC.57.1595}, the difference between $\rm R_3$ and our ratio would be less than 1\%. From our data, the measured ratio is \ratio. Sources of systematic uncertainty are the same as discussed earlier.  

\begin{figure}
\includegraphics[width=0.5\textwidth]{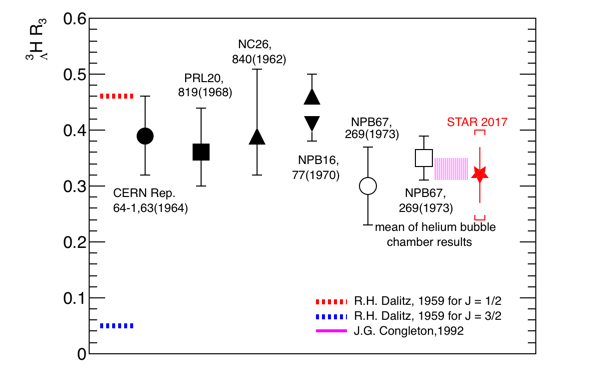}
\caption{(Color Online) A summary of worldwide \hla $R_3$ experimental measurements and theoretical calculations. The star marker represents the present analysis. }
\label{Fig.Wdata_R3}
\end{figure}

Fig.~\ref{Fig.Wdata_R3} summarizes previous measurements of this decay branching ratio in the literature. The present result is close to the combined measurement from helium bubble chamber experiments and is consistent with the average value of 0.35 $\pm$ 0.04 based on early measurements in helium bubble chambers. 

The branching fraction for the various decay modes of a hypernucleus will generally depend on both the spin of the hypernucleus and the nature of the $\Lambda-$decay interaction \cite{1959.PR.116.1312}. From the calculations in Ref.~\cite{1959.PR.116.1312}, our measurement lies within 1.5$\sigma$ of the calculated value under the assumption $J(\mathrm{^{3}_{\Lambda}H})$ = $\frac{1}{2}$ but 3$\sigma$ away under the assumption $J(\mathrm{^{3}_{\Lambda}H})$ = $\frac{3}{2}$. Furthermore, the $J(\mathrm{^{3}_{\Lambda}H})$ = $\frac{1}{2}$ assignment is consistent with the calculation $R_3 = 0.33 \pm 0.02$, where the \hla wave function was found in the context of a $\Lambda d$ two-body picture of the three-body bound state \cite{1992.JPG.18.339}.

In summary, we have presented a \hla lifetime measurement of $\tau =$ \lifetime ps as well as a measurement of the ratio of two of the \hla decay modes. A short \hla lifetime compared with that of the free $\Lambda$ ($\tau_{(\mathrm{^{3}_{\Lambda}H})}/\tau_{(\Lambda)}=0.54^{+0.09}_{-0.08}(\rm stat.)$) is reported, which may indicate that the $\Lambda$-$N$ interaction in \hla is stronger than previously believed. In addition, our measurement indicates that \hla more likely has an assignment of $J(\mathrm{^{3}_{\Lambda}H})$ = $\frac{1}{2}$ than $J(\mathrm{^{3}_{\Lambda}H})$ = $\frac{3}{2}$. Our results challenge the conventional understanding of the \hla as a weakly-bound $\Lambda d$ system, with more theoretical progress needed to understand the structure of this and other light hypernuclei. The STAR experiment will collect large datasets for Au+Au collisions over a range of beam energies during 2019-20, which will further reduce the uncertainty on the \hla lifetime and will likely provide new insight into the structure of the $\mathrm{^{3}_{\Lambda}H}$.

We thank the RHIC Operations Group and RCF at BNL, the NERSC Center at LBNL, and the Open Science Grid consortium for providing resources and support. This work was supported in part by the Office of Nuclear Physics within the U.S. DOE Office of Science, the U.S. National Science Foundation, the Ministry of Education and Science of the Russian Federation, National Natural Science Foundation of China, Chinese Academy of Science, the Ministry of Science and Technology of China (973 Program No. 2014CB845400) and the Chinese Ministry of Education, the National Research Foundation of Korea, GA and MSMT of the Czech Republic, Department of Atomic Energy and Department of Science and Technology of the Government of India; the National Science Centre of Poland, National Research Foundation, the Ministry of Science, Education and Sports of the Republic of Croatia, RosAtom of Russia and German Bundesministerium fur Bildung, Wissenschaft, Forschung and Technologie (BMBF) and the Helmholtz Association.

\bibliography{paper_submitted}
\end{document}